\documentclass[aps,prb,preprint,groupedaddress]{revtex4-1}

\usepackage[pdftex]{graphicx}
\usepackage{multirow}
\usepackage{amsmath}
\usepackage{units}

\providecommand{\ADDsG}{\ensuremath{\text{A}_{2D}/\text{A}_{G}}}
\providecommand{\celsius}{\ensuremath{^{\circ}\text{C}}}

\begin{document}


\title{Reversible optical doping of graphene}


\author{A. Tiberj$^{1,2}$}
\author{M. Rubio-Roy$^3$}
\author{M. Paillet$^{1,2}$}
\author{J.-R. Huntzinger$^{1,2}$}
\author{P. Landois$^{1,2}$}
\author{M. Mikolasek$^{1,2}$}
\author{S. Contreras$^{1,2}$}
\author{J.-L. Sauvajol$^{1,2}$}
\author{E. Dujardin$^3$}
\author{A.-A. Zahab$^{1,2}$}

\affiliation{$^1$Universit\'e Montpellier 2, Laboratoire Charles Coulomb UMR 5221, F-34095, Montpellier, France}
\affiliation{$^2$CNRS, Laboratoire Charles Coulomb UMR 5221, F-34095, Montpellier, France}
\affiliation{$^3$CEMES-CNRS, Universit\'e de Toulouse, 29 rue Jeanne Marvig, Toulouse 31055, France}

\author{}
\affiliation{}


\date{\today}

\begin{abstract}
The ultimate surface exposure provided by graphene monolayer makes it the ideal sensor platform but also exposes its intrinsic properties to any environmental perturbations. In this work, we demonstrate that the charge carrier density of graphene exfoliated on a SiO$_2$/Si substrate can be finely and reversibly tuned between electron and hole doping with visible photons. This photo-induced doping happens under moderate laser power conditions but is significantly affected by the substrate cleaning method. In particular, it is found to require hydrophilic substrates and to vanish in suspended graphene. These findings suggest that optically gated graphene devices operating with a sub-second time scale can be envisioned but also that Raman spectroscopy is not always as non-invasive as generally assumed.
\end{abstract}


\maketitle


Optical and electronic properties of graphene can be modulated by continuously tuning the charge carrier density using electrostatic gating\cite{Pisana:2007uq,PhysRevLett.98.166802,Das:2008yq,PhysRevB.79.155417,Chen:2011zr,PhysRevB.86.195434}, electrochemical doping \cite{doi:10.1021/nn1010914} or charge transfer by adsorption of molecular species\cite{doi:10.1021/nl902362q,ADFM:ADFM201000641,PhysRevB.81.125421,PhysRevB.82.075422,PhysRevB.82.245423,doi:10.1021/ja110939a,doi:10.1080/17458080.2010.529174,doi:10.1021/nn201368g,doi:10.1021/jz201273r,ADFM:ADFM201100401,PhysRevB.83.241401,PhysRevB.84.241404,C2JM32716C,doi:10.1021/nn300252a,Peimyoo2012201,doi:10.1021/nn3048878}. Besides electronic transport, the doping of graphene has a marked effect on the fundamental electron-phonon interactions, like the breakdown of the adiabatic Born-Oppenheimer approximation\cite{Pisana:2007uq,PhysRevLett.98.166802,Das:2008yq,PhysRevB.79.155417}, the interplay between adiabatic and non-adiabatic effects\cite{doi:10.1021/nn3048878} or the interference between all the quantum pathways involved in inelastic light scattering\cite{Chen:2011zr}. These effects have been conveniently investigated by Raman spectroscopy, which is also sensitive to the number of layers\cite{PhysRevLett.97.187401}, their stacking ordering\cite{PhysRevB.79.195417,doi:10.1021/nl1032827,doi:10.1021/nl301137k,PhysRevLett.108.246103}, the nature and density of defects\cite{PhysRevB.82.125429,PhysRevB.84.035433,doi:10.1021/nl300901a} and the in-plane strain variations\cite{PhysRevB.79.205433,Lee:2012uq}. As a consequence, Raman spectroscopy of active modes in graphene, like the G and 2D bands is being considered as a high-throughput technique to characterize graphene and to probe the inelastic light scattering phenomena. Moreover, several groups\cite{PhysRevLett.97.187401,doi:10.1021/nl8014439,doi:10.1021/nl8031444} have reported that no significant spectral changes were observed by collecting Raman spectra of graphene exfoliated on SiO$_2$/Si substrate, with a laser power P$_{\text{laser}}$ ranging from 0.04 to 4~mW. Therefore, Raman spectroscopy is considered as non-invasive when performed with P$_{\text{laser}}$ in the mW range.

Yet, the non-invasive character of Raman spectroscopy has been questioned by the reports of possible photo-induced effects and some authors recommended to use more cautious experimental conditions (laser power limited to 70~$\mu \text{W}$\cite{doi:10.1021/nl1029607} or Ar annealing\cite{doi:10.1021/nl8014439}). Laser irradiation has even been shown to induce irreversible damages of graphene\cite{krauss09,doi:10.1021/jp305823u}. Interestingly, we have found that the charge carrier density of exfoliated graphene lying on a hydrophilic SiO$_2$/Si substrate can be finely and reversibly tuned optically with visible photons. The influence of the laser power and the surface chemistry of the SiO$_2$/Si substrate on optical doping of exfoliated graphene in air is examined. Our result also implies that Raman spectroscopy of graphene performed with usual incident laser power density (around mW/$\mu \text{m}^2$) layers can be invasive as it directly alters the charge carrier density.

\section*{Results}
\begin{figure}[t]
\includegraphics[width=17 cm]{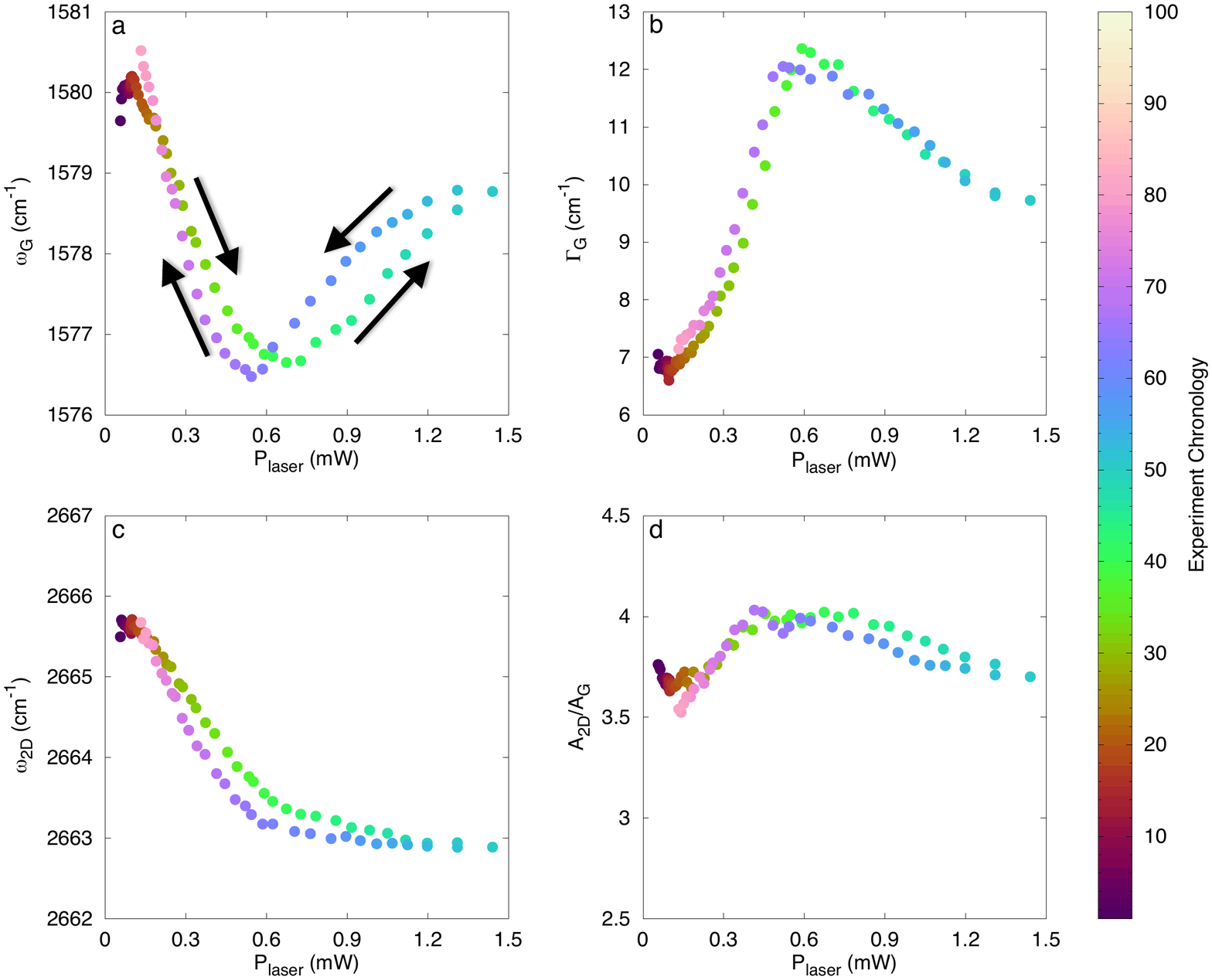}
\caption{\label{exfoliated_doping} \textbf{Reversible evolution of Raman spectra as a function of the incident laser power P$_{\text{laser}}$ for the graphene flake F1.} a) the G band position $\omega_G$, b) the G band FWHM $\Gamma_G$, c) the 2D band position $\omega_{2D}$ and d) the integrated intensities ratio \ADDsG\ . The graphene flake F1 is exfoliated on a hydrophilic substrate. P$_{\text{laser}}$ is increased from $0.05~\text{mW}$ up to 1.5 mW and decreased back to $0.05~\text{mW}$ as shown by the arrows in (a). The color code of each point corresponds to the chronological order in which the measurements have been carried out as depicted on the right hand side color bar.} 
\end{figure}

In a first series of experiments, Raman spectra of exfoliated graphene are collected as a function of the incident laser power, P$_{\text{laser}}$, at a fixed location. The results obtained for three different samples are compared. F1 was exfoliated on a hydrophilic SiO$_2$/Si substrate, F2 on a less hydrophilic substrate and F3 was suspended over a trench etched into the substrate (see methods). The quality of the samples is preserved during the whole experiments, as evidenced by the absence of the D band (see Supplementary Fig.~\ref{OM_spectra}). 
All Raman spectra are fitted by Lorentzian functions to extract the position of the G band ($\omega_G$), the full width at half maximum (FWHM) of the G band ($\Gamma_G$), the 2D band position ($\omega_{2D}$) and the ratio between the integrated intensities of the 2D and G bands (\ADDsG). For the sample F1, P$_{\text{laser}}$ is increased from $0.05~\text{mW}$ up to 1.5 mW and decreased back to $0.05~\text{mW}$, as shown in Figure~\ref{exfoliated_doping}. The G band at 1580 cm$^{-1}$ for $\text{P}_{\text{laser}} = 0.05~\text{mW}$ downshifts to reach a minimum value 1576.5~cm$^{-1}$ for $\text{P}_{\text{laser}} = 0.5-0.6~\text{mW}$. With the further increase of P$_{\text{laser}}$, it  upshifts up to 1579~cm$^{-1}$. Correspondingly, $\Gamma_G$, rises from 7~cm$^{-1}$ up to 12~cm$^{-1}$ for $\text{P}_{\text{laser}} = 0.5-0.6~\text{mW}$ and then decreases to 10~cm$^{-1}$ (Fig.~\ref{exfoliated_doping}b). In the same conditions, the 2D band continuously downshifts from 2666~cm$^{-1}$ down to 2663~cm$^{-1}$ (Fig.~\ref{exfoliated_doping}c). The \ADDsG\ ratio displays a maximum for P$_{\text{laser}}$ around $0.5-0.6~\text{mW}$ (Fig.~\ref{exfoliated_doping}d). 
Comparing the observed concomitant evolution of $\omega_G$, $\omega_{2D}$, $\Gamma_G$ and \ADDsG\ with electrostatic gating results\cite{Pisana:2007uq,Das:2008yq,PhysRevB.79.155417,PhysRevLett.98.166802} indicates that the graphene doping is modified upon light irradiation. In particular, the coincidence of the maxima of $\Gamma_G$ and of \ADDsG\ with the minimum of $\omega_G$ is a signature of neutral graphene. The doping type is determined from the evolution of $\omega_{2D}$:\cite{Das:2008yq,PhysRevB.79.155417} larger upshift for P-doping (below $0.5-0.6~\text{mW}$), than for low-level N-doping (above $0.5-0.6~\text{mW}$). Interestingly, the changes are observed to be mostly reversible when P$_{\text{laser}}$ is decreased back to its initial value (Fig.~\ref{exfoliated_doping}) so that the doping of graphene deposited on a hydrophilic SiO$_2$/Si substrate can be continuously and reversibly tuned from the initial P-doping to quasi-neutral and eventually N-doping with moderate laser power. From the shifts and widths of the G band\cite{Das:2008yq, PhysRevB.79.155417} the inital doping of F1 was estimated to be $\text{p}\approx4\times 10^{12}~\text{cm}^{-2}$  at $0.05~\text{mW}$. For $\text{P}_{\text{laser}} = 1.44~\text{mW}$, the carrier density was estimated to be $\text{n} \approx 3\times 10^{12}~\text{cm}^{-2}$.

\begin{figure}[t]
\includegraphics[width=17 cm]{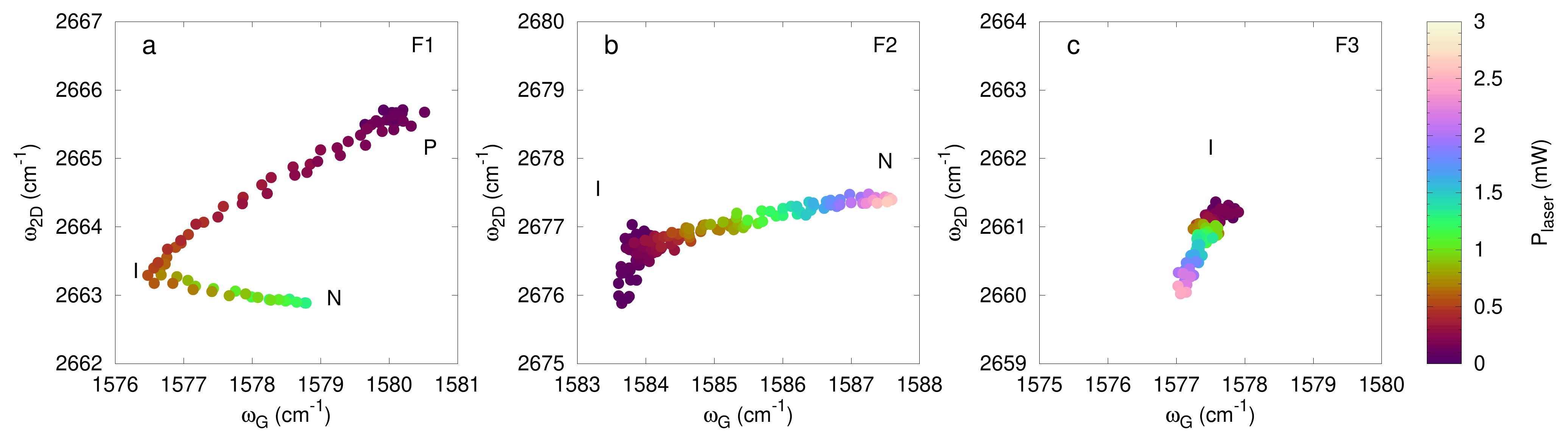}
\caption{\label{surface_treatment} \textbf{Comparison of the relative evolutions of the 2D band position ($\omega_{2D}$) versus the G band position ($\omega_G$) as a function of P$_{\text{laser}}$ for three graphene flakes (F1, F2 and F3)}. F2 was exfoliated on a less hydrophilic SiO$_2$/Si substrate than F1. F3 was suspended over a trench etched into the substrate. 
The color code of each point corresponds to the incident laser power P$_{\text{laser}}$ as displayed on the right hand side color bar.
a) F1 is P-doped at low P$_{\text{laser}}$, it becomes quasi-neutral around 0.5 mW and N-doped for higher P$_{\text{laser}}$. b) F2 is initially quasi-neutral and becomes N-doped with the increasing P$_{\text{laser}}$. c) The suspended graphene flake, F3, is neutral and stays neutral with the increasing P$_{\text{laser}}$. The measured shifts for F3 are only due to laser heating effects.
The different G band and 2D band positions ($\omega_G$, $\omega_{2D}$) of the different graphene samples (F1, F2, F3) in their quasi-neutral state are attributed to strain fluctuations from one sample to the other.
}
\end{figure}

We now examine the influence of the underlying substrate by comparing these results to the ones obtained on sample F2 deposited on a less hydrophilic substrate and to sample F3 suspended over a trench etched into the substrate. The relative variations of the 2D and G band positions as a function of P$_{\text{laser}}$ are collected in Figure~\ref{surface_treatment}. 
This particular representation disentangles doping and strain effects\cite{Lee:2012uq}. We clearly observe three different behaviors in the plots presented on Figure~\ref{surface_treatment}. When the doping of graphene is continuously tuned from P-type to N-type (Fig.~\ref{exfoliated_doping}), the 2D versus G positions presents the particular v-shape shown in Figure~\ref{surface_treatment}a. 
The G band consistently upshifts with an increasing charge carrier density (for both electrons and holes). 
The 2D band downshifts monotonically as the doping varies from P to N. Figure~\ref{surface_treatment}b displays the 2D vs G behavior for a flake lying on a less hydrophilic substrate. The G and 2D bands continuously upshift as P$_{\text{laser}}$ increases. The small upshift of the 2D band is expected for low level N-type doping \cite{Das:2008yq, PhysRevB.79.155417,doi:10.1021/nn3048878}.
Here, $\Gamma_G$ is maximum for the lowest incident power and continuously decreases with the increasing laser power, and so does the \ADDsG\ ratio (see Supplementary Fig.~\ref{annealed_doping}). Considered together, these variations indicate that this supported graphene is close to neutral for the lowest P$_{\text{laser}}$ and becomes N-doped as P$_{\text{laser}}$ is increased (for $\text{P}_{\text{laser}} = 2.6~\text{mW}$, $\text{n}\approx 4-5\times 10^{12}~\text{cm}^{-2}$ from refs.~\onlinecite{Das:2008yq, PhysRevB.79.155417}). Once again, these doping variations are reversible when P$_{\text{laser}}$ is decreased back. For the suspended graphene flake, F3, the high values of $\Gamma_G$ (14~cm$^{-1}$) and \ADDsG\ ratio (9.5) remains constant through the entire power sweep (Supplementary Fig.~\ref{suspended_doping}). This indicates that F3 remains neutral as the laser power increases in agreement with S. Berciaud~\textit{et al.}\cite{doi:10.1021/nl8031444} Thus, the small shifts of $\omega_G$ and $\omega_{2D}$ (Fig.~\ref{surface_treatment}c) cannot be ascribed to any doping level variation.

\begin{figure}[t]
\includegraphics[width=16 cm]{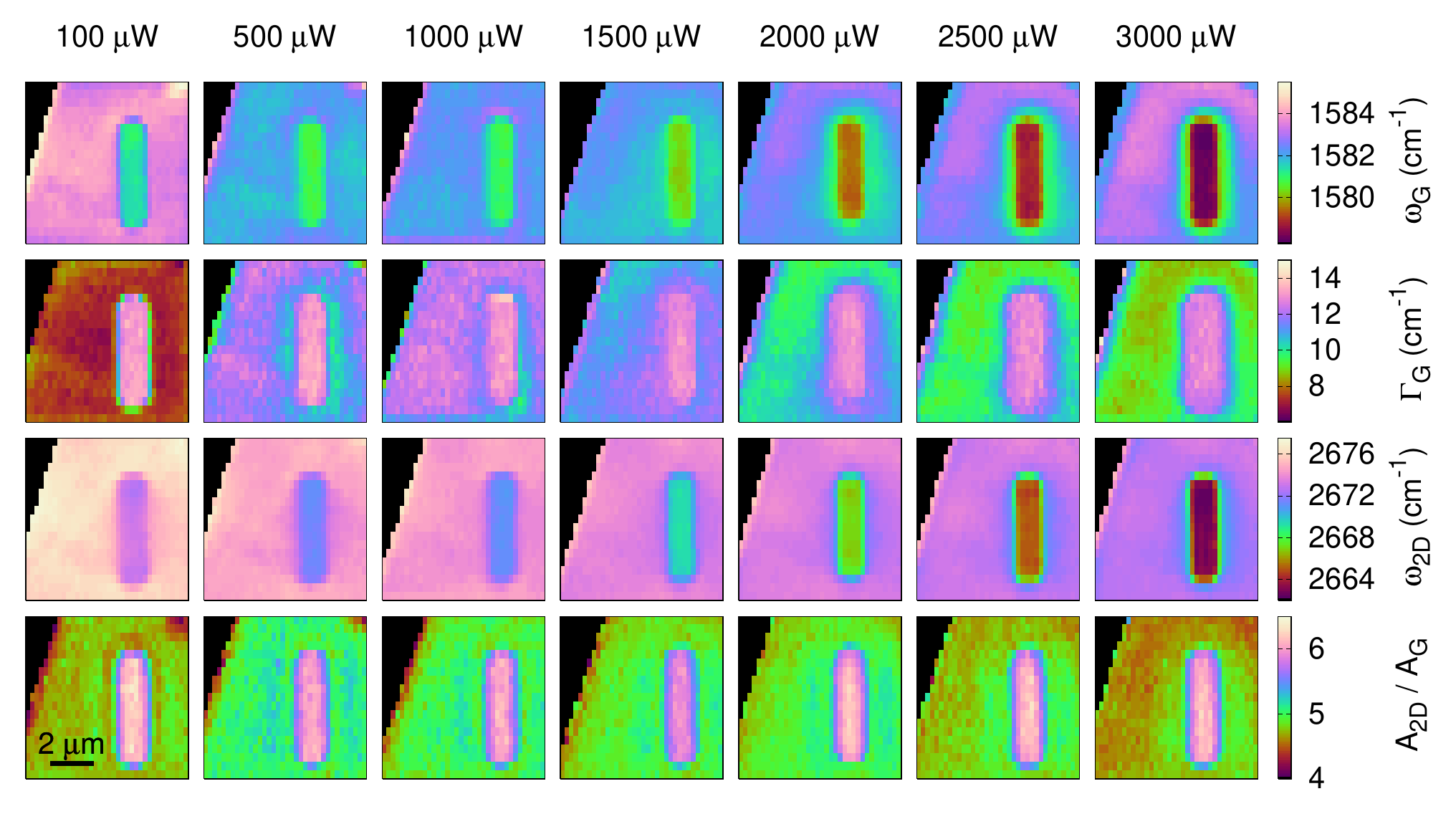}
\caption{\label{surface_treatment_map} \textbf{Raman maps measured for 7 different laser powers on sample F4}. The rows correspond from top to bottom to the G band position $\omega_G$, the G band FWHM $\Gamma_G$, the 2D band position $\omega_{2D}$, and the integrated intensities ratio \ADDsG. 
The graphene flake is covering an etched pool of the substrate that has a capsule shape and is visible on the right hand side of each map. The black upper left corners correspond to bare SiO$_2$/Si surfaces.
}
\end{figure}

We can now wonder how these laser induced doping variations are uniform across the graphene surface. To this end, Raman mapping was performed at different laser powers on graphene exfoliated onto a hydrophilic substrate and partially suspended over a pre-patterned trench (sample F4). Maps of the $\omega_G$, $\Gamma_G$, $\omega_{2D}$ and \ADDsG\  are displayed on Figure~\ref{surface_treatment_map} for seven P$_{\text{laser}}$ ranging from $100~\mu \text{W}$ to 3~mW. The constant position of the G and 2D bands across the whole supported area of the sample indicates that the strain and doping are homogeneous, therefore allowing to single out the influence of optical doping by the varying laser power. When the laser power is increased from $100~\mu \text{W}$ to 1.5~mW, the G band, initially at 1584~cm$^{-1}$ ($\Gamma_G$ = 7.5~cm$^{-1}$) shifts down to 1582~cm$^{-1}$ ($\Gamma_G$ = 12~cm$^{-1}$). Further increase of the laser power inverts the effect and the G band upshifts back to 1582.5~cm$^{-1}$ ($\Gamma_G$ decreases down to 9.5~cm$^{-1}$). For the same power sweep, the 2D band continuously downshifts from 2676 to 2672~cm$^{-1}$. Similarly, the \ADDsG\  ratio increases from 4.7 for $\text{P}_{\text{laser}} = 100~\mu \text{W}$, to 5.1 for P$_{\text{laser}}$ between 0.5 and 1.5~mW, and then decreases back to 4.7 for higher laser powers. These observations confirm that graphene supported on hydrophilic SiO$_2$ evolves continuously from P-type to N-type doped graphene when the laser power is increased.

Interestingly, the irradiation of the suspended graphene yields a very different behavior. Indeed, Figure~\ref{surface_treatment_map} shows that both G and 2D bands downshifts from 1581.5 to 1578.5~cm$^{-1}$ and from 2673 to 2663.5~cm$^{-1}$ respectively as the laser power is increased. However, $\Gamma_G$ and the \ADDsG\  ratio remain constant (13 cm$^{-1}$ and 5.9 respectively). Therefore the doping level remains unchanged through the entire power sweep and the downshifts of the 2D and G bands are ascribed to classical laser-induced heating effects that are also observed on samples F1 and F3 (see Supplementary Fig.~\ref{heating}).

A careful examination of the signals recorded on the graphene edges and on the trench contour reveals that the optical doping is less effective in these specific locations. In particular, the doping never reaches N-type even for the highest P$_{\text{laser}}$. We attribute this local effect to the supplementary P-doping near graphene edges which has already been observed by Raman spectroscopy\cite{doi:10.1021/nn200550g,doi:10.1021/nl8032697}.

\section*{Discussion}
The absence of doping upon irradiation of suspended graphene clearly shows that the substrate contributes to the mechanism involved in the optical doping of graphene. We thus investigated the influence of substrate cleaning procedures on the efficiency of the optical doping. For highly hydroxylated (hydrophilic) substrates freshly cleaned by O$_2$ plasma and/or Piranha treatments prior to graphene deposition, we observe the ambipolar behavior with a doping evolving from P-type to N-type upon increasing laser power. Our results confirm that graphene exfoliated on a hydrophilic substrate is P-doped\cite{PhysRevB.80.233407,doi:10.1021/nl103015w} in contrast to the quasi-intrinsic state of suspended graphene\cite{doi:10.1021/nl8031444}. Across the different samples studied here, we found that the P$_{\text{laser}}$ for which graphene is neutralized falls in the range 0.5-1.5 mW. We anticipate that these values can change with the laser wavelength and the thickness of the SiO$_2$ layer. The density of hydroxyl groups can be reduced by performing a thermal annealing of the samples under argon before graphene deposition. In this case, graphene was observed to be initially quasi-neutral and N-type doping was obtained upon illuminating the sample with increasing P$_{\text{laser}}$. Finally, when graphene is transferred onto substrates used as-received without any cleaning, the doping of some flakes remain constant for all the entire P$_{\text{laser}}$ range (not shown). Therefore, the absence of any significant spectral changes previously observed\cite{PhysRevLett.97.187401} might be accounted for by the deposition of graphene on as-received substrates. Finally, it should be pointed out that optical doping is not specific to graphene micromechanically exfoliated onto SiO$_2$/Si substrates but similarly occurs for graphene deposited on standard glass (not shown)

It thus appears that the starting value of the graphene doping level is directly linked to the substrate preparation. Cleaning procedures that render the substrate hydrophilic tend to result in more P-doped supported graphene, in agreement with previous studies\cite{doi:10.1021/nl903162a, wang10a, nagashio:024513, lee11, doi:10.1021/nn3017603}, which also demonstrated the involvement of adsorbed water in the P-doping of graphene by atmospheric oxygen\cite{doi:10.1021/nl103015w,doi:10.1021/nl1029607,doi:10.1021/nn3017603,nagashio:024513,SMLL:SMLL201102468}. Irreversible or slow modulation of charge carrier density was demonstrated by changing the atmosphere\cite{doi:10.1021/nl1029607,doi:10.1021/nl103015w} or by illuminating graphene with UV or visible light\cite{,doi:10.1021/nl1029607,luo12,krauss09}. In the latter cases, the invoked mechanisms relied on UV photons or laser heating assisted removal of dopant molecules and more specifically oxygen derivatives. 
The characteristic times of such phenomena were found to be in the range of minutes \cite{doi:10.1021/nl1029607,luo12} or even hours\cite{krauss09}. Other authors probed the dynamics of charge transfer between the H$_2$O/O$_2$ redox couple and graphene by electrical measurements alone or combined with Raman spectroscopy. 
It was shown that the equilibrium is established after minutes \cite{SMLL:SMLL201102468} or hours\cite{doi:10.1021/nl103015w}. 

Our study shows that the charge carrier density can be conveniently tuned by adjusting the incident laser power even without a gating electrode. This effect does not involve the chemical modification of graphene since no D band emerges upon extended irradiation. By contrast with previous reports\cite{doi:10.1021/nl1029607,luo12,krauss09}, this laser-induced doping is reversible with a characteristic time that was preliminarily evaluated to be less than 1~s, \textit{i.e.} orders of magnitude faster. This suggests that this phenomenon, although related to similar environmental effects, involves a different mechanism. O$_2$ and H$_2$O are playing a key role as illustrated by the weak dependency of $\omega_G$ on P$_\text{laser}$ found after adsorbates removal by Ar annealing at 150\celsius\ of graphene lying on O$_2$ plasma treated SiO$_2$/Si substrates\cite{doi:10.1021/nl8014439}. Considering the H$_2$O/O$_2$ redox couple or O$_2^-$ superoxide anion, it has been suggested that optically excited graphene transfers ``hot" electrons\cite{doi:10.1021/nl1029607}. This would lead to an increase of hole doping in graphene upon visible light exposure, contrary to our observations. This suggests that other electrochemical reactions kinetically more favorable than the one considered previously might be involved.
Noteworthy, graphene lying on SiO$_2$/Si exposed to H$_2$O alone, with only traces of O$_2$, has been shown to be N-doped\cite{doi:10.1021/nl103015w}. Similarly, water significantly reduces hole doping of graphene deposited on mica\cite{shim12}. The elucidation of the laser assisted charge carrier density tuning mechanism is beyond the scope of this paper and deserves further investigation.

In conclusion, we have shown that a low power visible laser light can be used to reversibly tune the charge carrier density of graphene lying on a substrate. This effect is highly sensitive to the substrate hydrophilicity and completely suppressed in suspended graphene. The continuous tuning of the doping in graphene from P-type to N-type has been achieved on O$_2$ plasma treated SiO$_2$/Si substrates. The observed sub-second dynamics of the optical gating phenomenon points to a new underlying mechanism that remains to be elucidated.

One technical implication of our study for the entire scientific community using Raman spectroscopy of graphene as a routine characterization technique is that it should be considered as potentially invasive as far as electronic properties are concerned. In particular, the laser induced-modification of graphene doping could account for recent discrepancies between Raman and electrical transport measurements\cite{nagashio:024513}. It would be interesting to extend the present work to graphene on metals\cite{doi:10.1021/nn3017603} and on silicon carbide\cite{sidorov:113706} to assess how carefully Raman experiments on graphene must be performed.

On another hand, the ability to tune the charge carrier density with visible photons opens a wide set of opportunities to develop optically gated graphene electronic devices and a new approach to graphene optoelectronics. Finally, this effect should allow to study the interplay between graphene properties and the environment and to trigger laser-assisted functionalization of graphene leading to more advanced devices\cite{Wang:2012kx,doi:10.1021/cr3000412}.

\section*{methods}
\subsection*{Sample fabrication}
Four samples (F1 through F4) were prepared for this study. The four of them consisted on a $\unit[500]{\textrm{$\mu$m}}$ thick highly P-doped monocrystalline Si (100) substrate, with a thermal oxide layer of $\unit[290\pm5]{nm}$ (in F1 and F3, oxide grown in $\mathrm{O_2}$; in F2 and F4, oxide grown in $\mathrm{H_2O}$) and a square matrix of metallic marks every $\unit[200]{\textrm{$\mu$m}}$ to ease the identification of graphene flakes. Samples F3 and F4 were further processed by etching matrices of $\unit[5\times1]{\textrm{$\mu$m}}$ trenches with depths of $\unit[160]{nm}$ (F3) and $\unit[480]{nm}$ (F4) in between the metallic marks with a $\mathrm{CF_4}$ ICP-RIE process.

Cleaning of F1 and F3 was performed by sonication in organic solvents (N-methyl-2-pyrrolidone, acetone and isopropanol; VLSI quality), immersion in Piranha bath ($\mathrm{H_2SO_4/H_2O_2}$ 3:1) during $\unit[3]{hours}$ and $\mathrm{Ar/O_2}$ (3:1) plasma cleaning in a Fischione 1020 for $\unit[20]{min}$. On the other hand, F2 and F4 were cleaned by sonication in organic solvents and $\mathrm{O_2}$ plasma cleaning in a PVA TePLA 300 at $\unit[800]{W}$ for $\unit[15]{min}$. In addition, sample F2 was annealed during $\unit[1]{hour}$ at $\unit[400]{^\circ C}$ in a $\unit[300]{sccm}$ Ar flow at atmospheric pressure.

Finally, graphene was deposited on all samples by exfoliation from kish graphite with the scotch tape method\cite{Science_Novoselov_2004}.

\subsection*{Micro-Raman spectroscopy}
Raman spectra were recorded using an Acton spectrometer fitted with a Pylon CCD detector and a 600 grooves/mm grating (2.5 cm$^{-1}$ between each CCD pixel). 
The samples were excited with a 532 nm (2.33 eV) CW frequency doubled Nd:Yag laser through a x100 objective (N.A. 0.9). The gaussian laser spot FWHM is about 350 nm. Optimized focus conditions have been checked for each measurement.
The sample are mounted on a three-axis piezoelectric stage to ensure the precise positioning and focusing of the laser spot. The maps were recorded with a 0.2 $\mu$m step in X (0.4 $\mu$m in Y) to probe the suspended graphene independently with a minimum of 2 to 3 points in the middle of the pool.
The laser power was tuned with a variable neutral density filter controlled by a servomotor. The laser power was continuously measured by a calibrated photodiode put behind the beamsplitter. 
The whole experimental setup (spectrometer, piezoelectric stage, photodiodes, servomotor) were controlled by a dedicated and home-made Labview application.

\section*{Acknowledgments}
This work was supported by the french ANR (Grafonics project ANR-10-NANO-0004) and has been done in the framework of the GDRI GNT No 3217 “Graphene and Nanotubes: Sciences and Application”. 

\section*{Author contributions}
All authors contributed extensively to the work presented in this paper.

\section*{Supplementary Information}
Optical microscopy (OM) images of the samples F1, F3 and F4 are shown. Raman spectra of F1 for three different laser powers are compared to the typical Raman spectrum collected on the suspended graphene flake F3. Raman spectra evolutions of F2 and F3 samples as a function of P$_{\text{laser}}$ are displayed. Finally, evidences of laser-induced heating of the F1 and F3 samples are detailed.
\section*{Competing financial interests} 
The authors declare no competing financial interests. 

\newpage

\clearpage

\begin{center}
{\Large\textbf{Supplemental Information\\[0.3cm]}}
\textbf{Reversible optical doping of graphene\\}
A. Tiberj, M. Rubio-Roy, M. Paillet, J.-R. Huntzinger, P. Landois, M. Mikolasek, S.~Contreras, J.-L. Sauvajol, E. Dujardin, and A.-A. Zahab
\end{center}

\setcounter{figure}{0}
\makeatletter 
\renewcommand{\thefigure}{S\@arabic\c@figure}

\subsection*{Optical Microscopy images and Raman spectra of the graphene flakes}
\begin{figure}[h]
\begin{minipage}[l]{0.56\columnwidth}
\includegraphics[height=5.5cm]{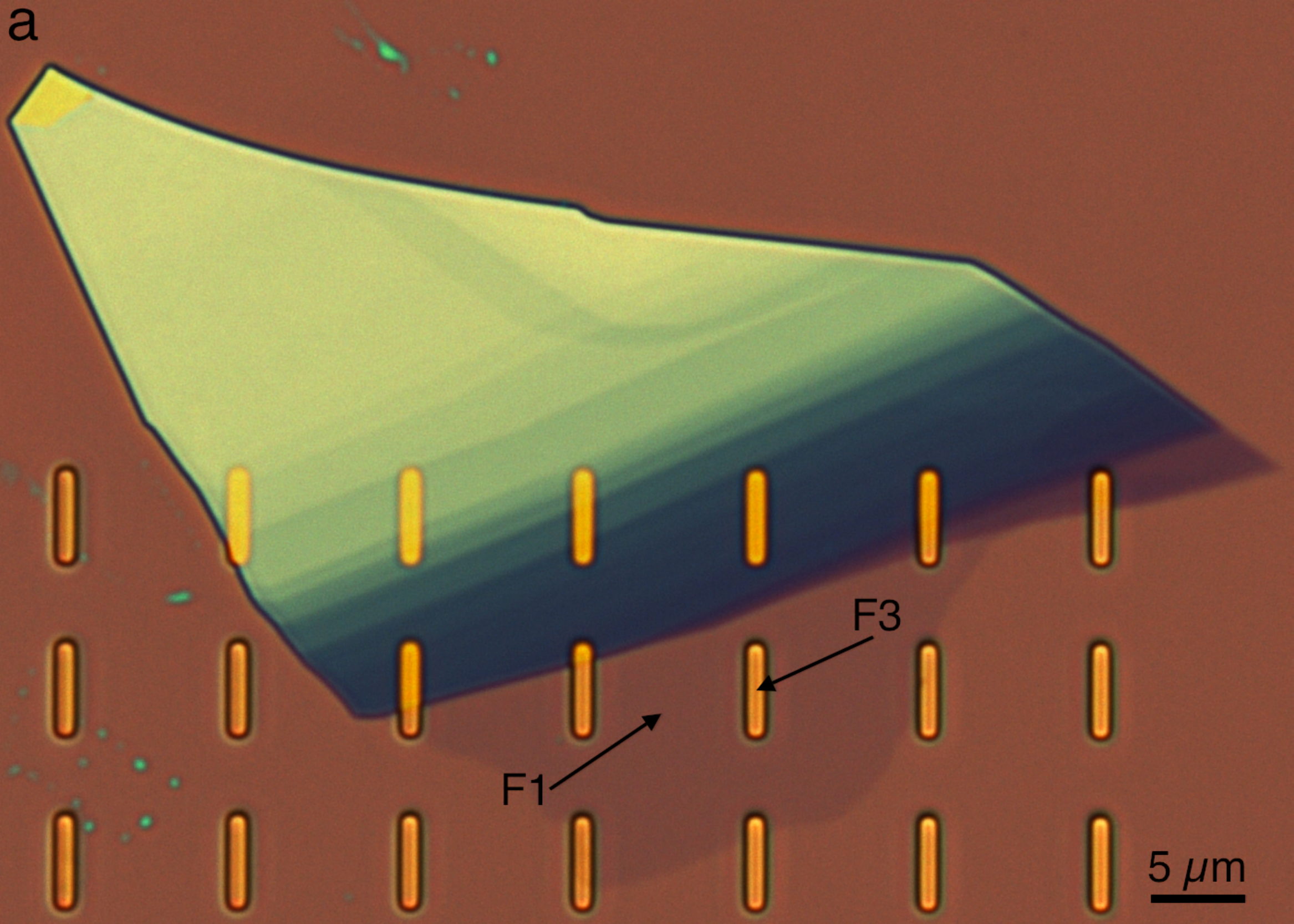}
\end{minipage}
\hspace{0cm}
\begin{minipage}[l]{0.24\columnwidth}
\includegraphics[height=5.5cm]{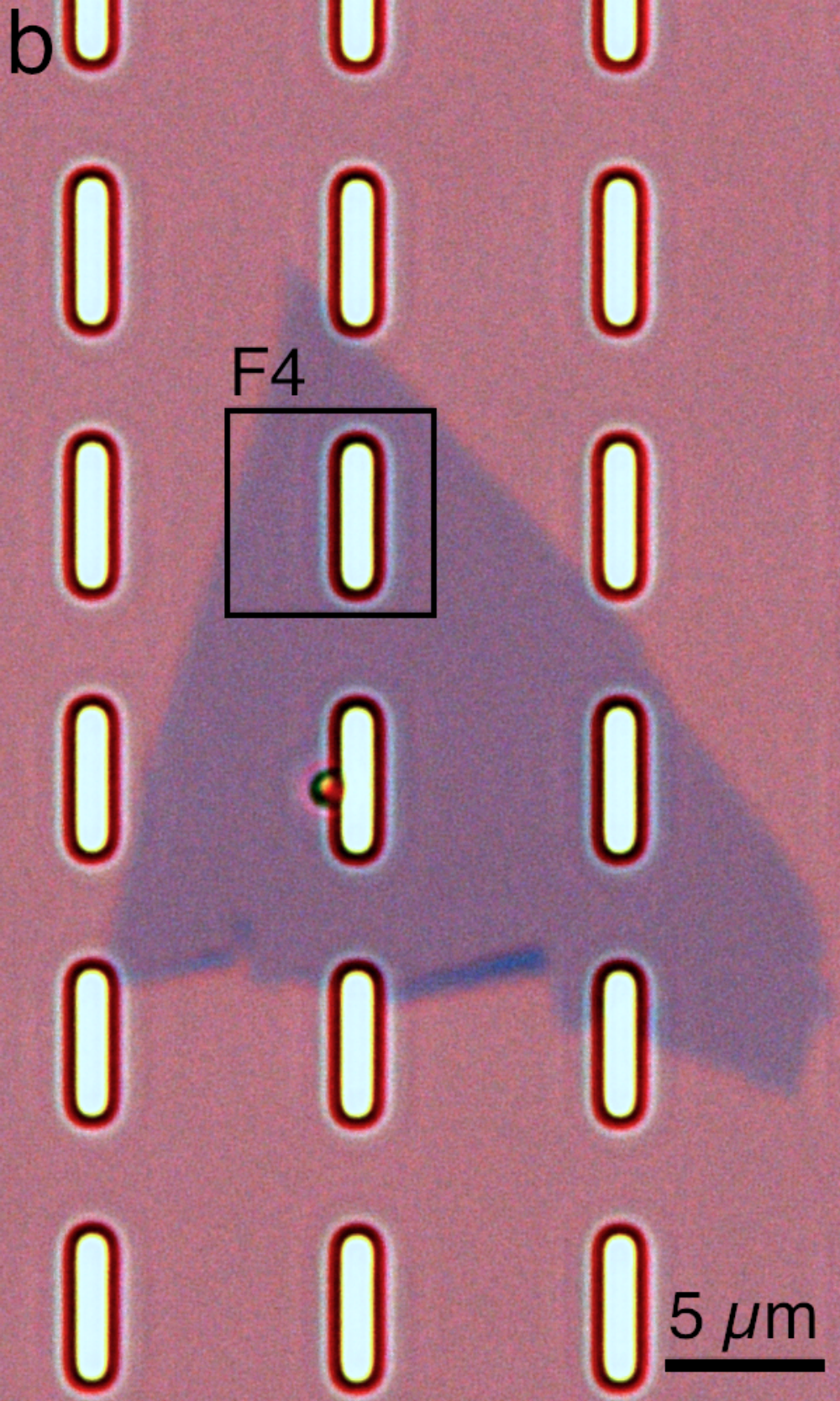}
\end{minipage}
\\[0.5cm]
\includegraphics[width=0.8\columnwidth]{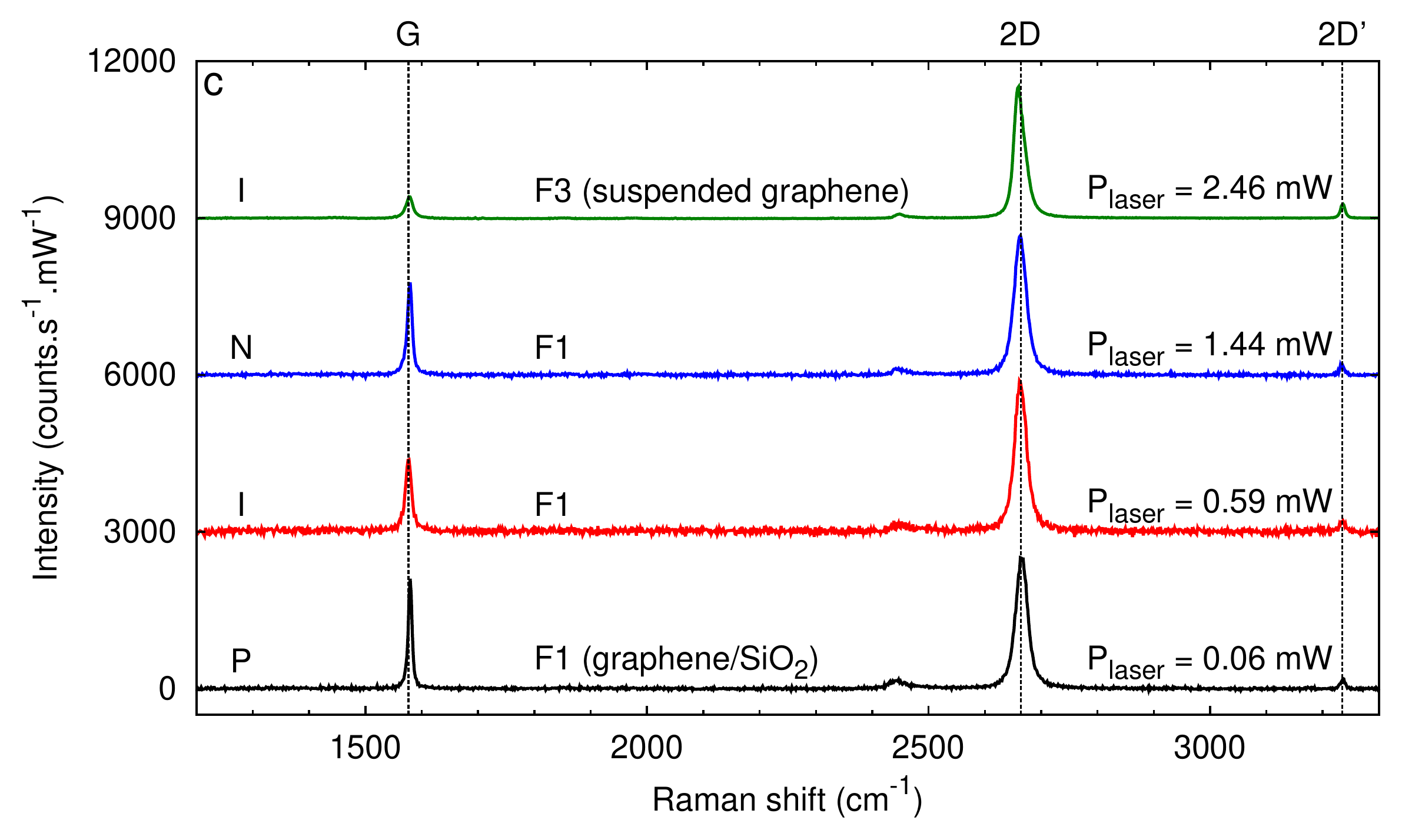}
\caption{\label{OM_spectra} \textbf{Optical Microscopy (OM) images and Raman spectra of the graphene flakes.} a) OM of F1 and F3 b) OM of F4. The black square corresponds to the area mapped in Fig.\ref{surface_treatment_map} c) The Raman spectra of F1 for 3 different laser powers are compared to the typical Raman spectrum collected on the suspended graphene flake F3.
 }
\end{figure}

\clearpage

\subsection*{Raman spectra evolutions of F2 and F3 flakes as a function of P$_{\text{laser}}$}
The graphene flake F2 is exfoliated on a less hydrophilic substrate than F1. P$_{\text{laser}}$ is increased from $0.05~\text{mW}$ up to 2.6 mW and decreased back to $0.05~\text{mW}$. The G and 2D bands continuously upshift as P$_{\text{laser}}$ increases. For this sample, $\Gamma_G$ is maximum for the minimum incident power and continuously decreases with the increase of P$_{\text{laser}}$, so does  \ADDsG. The moderate $\omega_{2D}$ upshift indicates low N-doping\cite{Das:2008yq,PhysRevB.79.155417}. All these variations indicate that this graphene flake is quasi-neutral for the lowest P$_{\text{laser}}$ and becomes N-doped as P$_{\text{laser}}$ is increased ($\text{n}\approx 4-5\times 10^{12}~\text{cm}^{-2}$ from refs.~\onlinecite{Das:2008yq, PhysRevB.79.155417}, for $\text{P}_{\text{laser}} = 2.6~\text{mW}$). \\

\begin{figure}[b]
\includegraphics[width=17 cm]{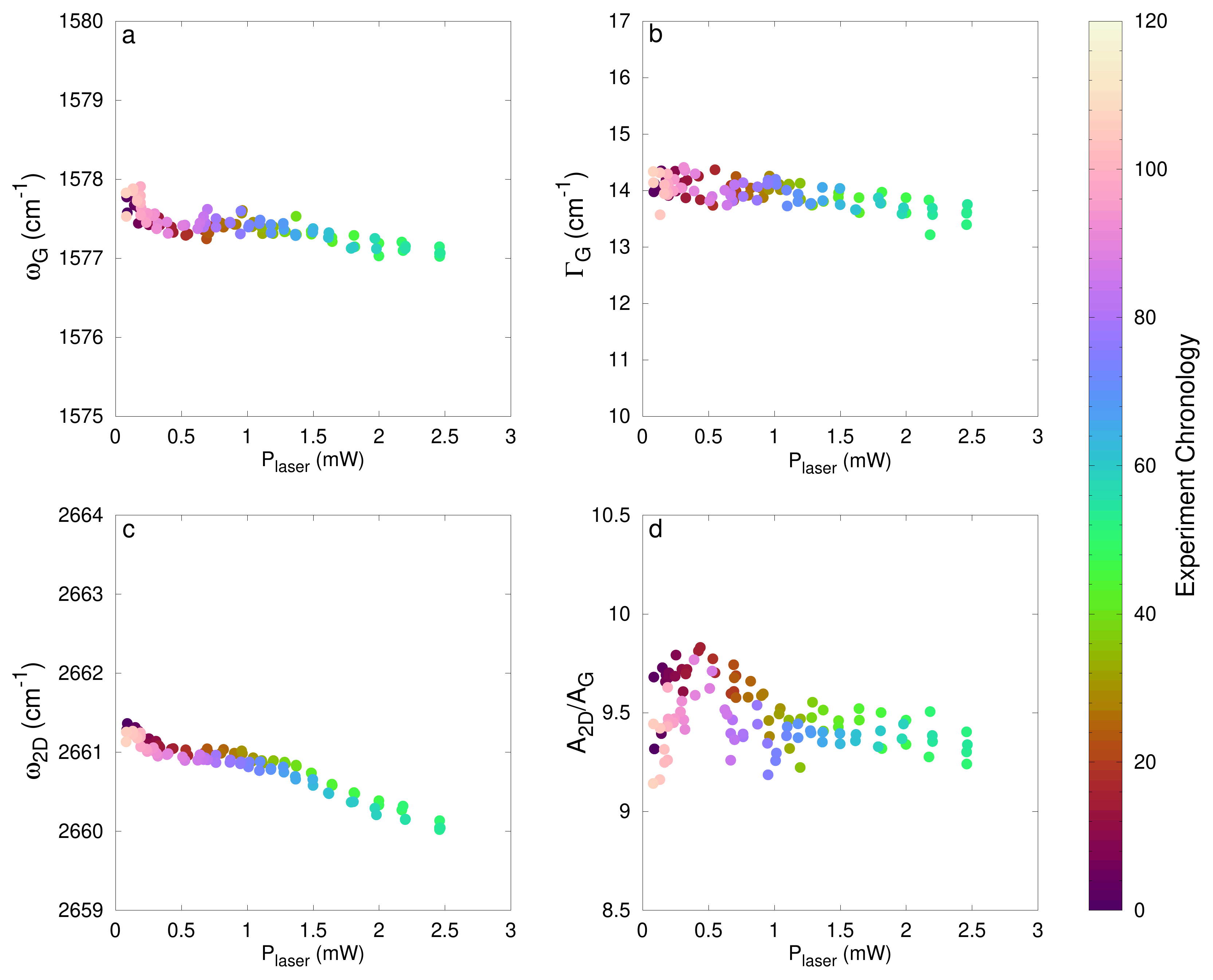}
\caption{\label{suspended_doping} \textbf{Reversible evolution of Raman spectra as a function of the incident laser power P$_{\text{laser}}$ for the graphene flake F3.} a) the G band position $\omega_G$, b) the G band FWHM $\Gamma_G$, c) the 2D band position $\omega_{2D}$ and d) the 2D/G integrated intensities ratio \ADDsG\  as a function of the incident laser power P$_{\text{laser}}$. F3 was suspended over a trench etched into the substrate. P$_{\text{laser}}$ is increased from $0.05~\text{mW}$ up to 2.5 mW and decreased back to $0.05~\text{mW}$. The color code of each point corresponds to the chronological order in which the measurements have been carried out as depicted on the right hand side color bar.
}
 
\end{figure}

For the suspended graphene flake, F3, the $\omega_G$ and $\omega_{2D}$ are almost constant as P$_{\text{laser}}$ increases (Fig.~\ref{suspended_doping}). This suspended flake is neutral and stays neutral as the P$_{\text{laser}}$ increases. It is confirmed by the constant  $\Gamma_G$ and the constant \ADDsG\  ratio (Fig.~\ref{suspended_doping}). The only small downshift of the G and 2D bands comes from the laser heating of the suspended flake (see the ``Heating effect'' section). \\

\begin{figure}[t]
\includegraphics[width=17 cm]{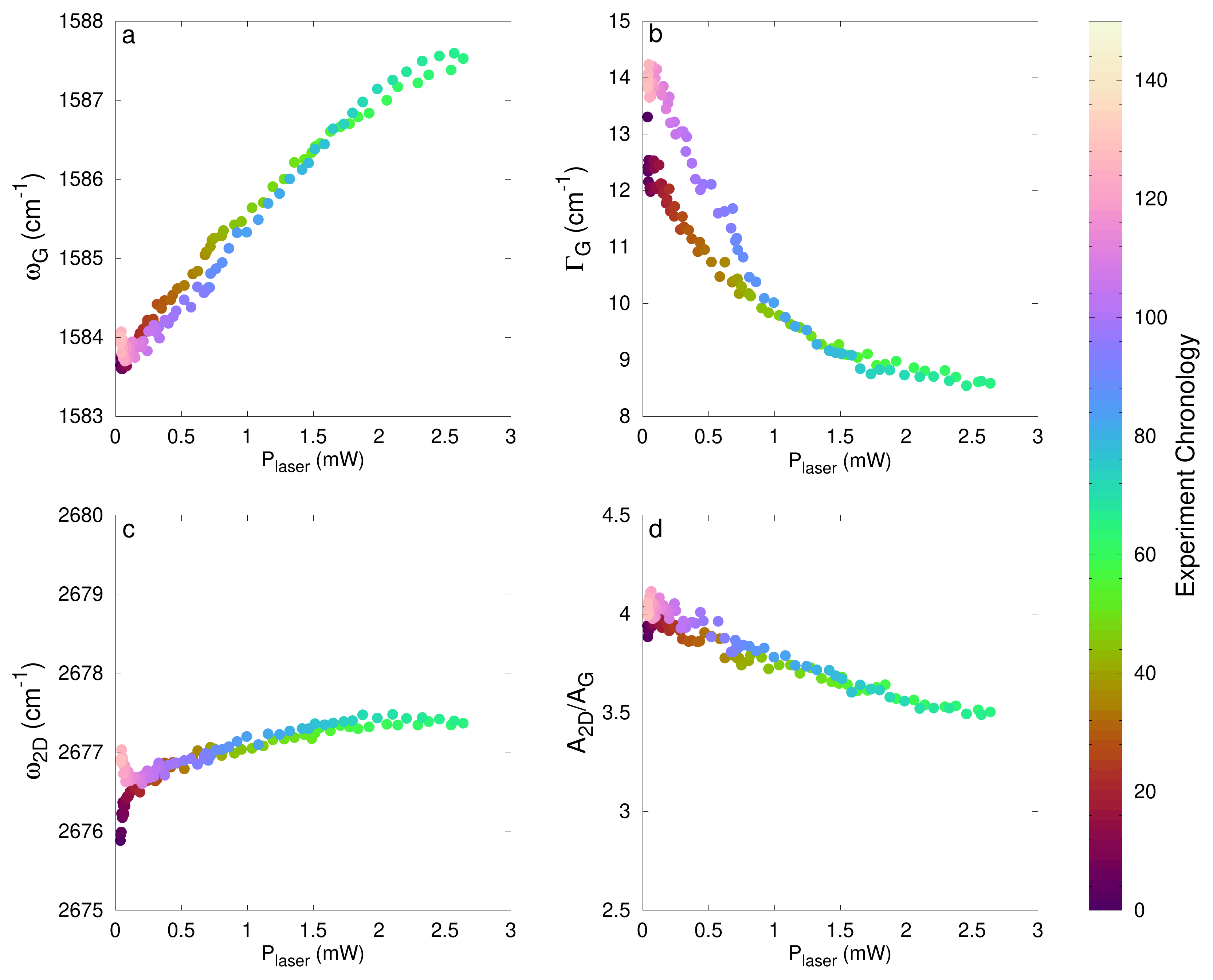}
\caption{\label{annealed_doping} \textbf{Reversible evolution of Raman spectra as a function of the incident laser power P$_{\text{laser}}$ for the graphene flake F2.} a) the G band position $\omega_G$, b) the G band FWHM $\Gamma_G$, c) the 2D band position $\omega_{2D}$ and d) the 2D/G integrated intensities ratio \ADDsG\  as a function of the incident laser power P$_{\text{laser}}$. F2 was exfoliated on a less hydrophilic substrate than F1. P$_{\text{laser}}$ is increased from $0.05~\text{mW}$ up to 2.6 mW and decreased back to $0.05~\text{mW}$. The color code of each point corresponds to the chronological order in which the measurements have been carried out as depicted on the right hand side color bar.
 }
\end{figure}

\clearpage
\subsection*{Heating effect}

\begin{figure}[t]
\includegraphics[width=17 cm]{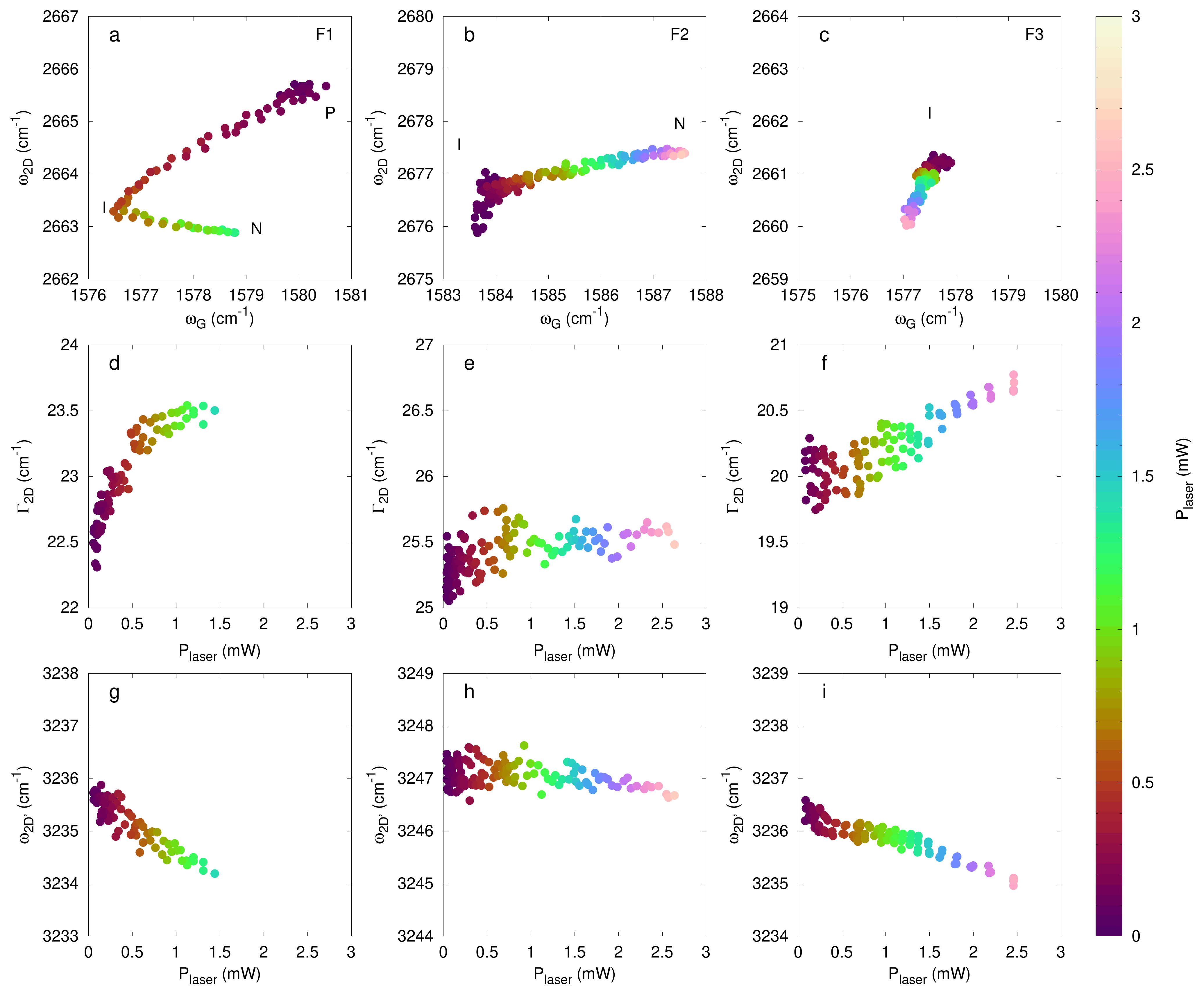}
\caption{\label{heating} \textbf{Raman fingerprint of a laser-induced heating effects.} Comparison of the relative evolutions of the 2D band position ($\omega_{2D}$) versus the G band position ($\omega_G$), the 2D band FWHM ($\Gamma_{2D}$) and the 2D'~band position ($\omega_{2D'}$) as a function of P$_{\text{laser}}$ for three graphene flakes (F1, F2 and F3). F2 was exfoliated on a SiO$_2$/Si substrate less hydrophilic than F1. F3 was suspended over a trench etched into the substrate. 
The color code of each point corresponds to the incident laser power P$_{\text{laser}}$ as displayed on the right hand side color bar. For the graphene flakes F1 and F3 the 2D band broadens and the 2D'~band downshifts as P$_{\text{laser}}$ increases. These variations are ascribed to a laser induced heating. In opposition, the graphene flake F2 does not exhibit laser power dependance of $\Gamma_{2D}$ and $\omega_{2D'}$ and is therefore not heated for this laser power range.
}
\end{figure}

Fig.~\ref{heating} displays how the 2D band FWHM ($\Gamma_{2D}$) and the 2D'~band position ($\omega_{2D'}$) depend on the incident laser power P$_{\text{laser}}$. For the graphene flakes F1 and F3 the 2D band broadens and the 2D'~band downshifts as P$_{\text{laser}}$ increases. F. Alzina~\textit{et al.} showed that the 2D'~band position does not depend on graphene doping level\cite{PhysRevB.82.075422}. Only strain and/or heating effects can induce such downshift.
Moreover, it has been also shown that the 2D band linewidth depends linearly on the temperature\cite{PhysRevB.83.125430}. Assuming that inhomogeneous broadening of the 2D band do not evolve with P$_{\text{laser}}$, we can deduce that this 2D band broadening and 2D' band downshift are due for F1 and F3 sample to heating effects. In opposition, the graphene flake F2 do not exhibit P$_{\text{laser}}$ dependance of $\Gamma_{2D}$ and $\omega_{2D'}$. It means that this flake is not heated for this P$_{\text{laser}}$ range, within the experimental precision.

\end{document}